\documentclass[prl,showpacs,amsmath,amssymb, twocolumn, 10pt]{revtex4}

\usepackage{amsthm}
\usepackage{dcolumn}
\usepackage{bm}
\usepackage{graphicx}

\begin{document}

\newtheorem{corollary}{Corollary}
\newtheorem{definition}{Definition}
\newtheorem{example}{Example}
\newtheorem{lemma}{Lemma}
\newtheorem{proposition}{Proposition}
\newtheorem{theorem}{Theorem}
\newtheorem{fact}{Fact}
\newtheorem{property}{Property}
\newcommand{\bra}[1]{\langle #1|}
\newcommand{\ket}[1]{|#1\rangle}
\newcommand{\braket}[3]{\langle #1|#2|#3\rangle}
\newcommand{\ip}[2]{\langle #1|#2\rangle}
\newcommand{\op}[2]{|#1\rangle \langle #2|}

\newcommand{\tr}{{\rm tr}}
\newcommand {\E } {{\mathcal{E}}}
\newcommand {\F } {{\mathcal{F}}}
\newcommand {\diag } {{\rm diag}}

\title{Entanglement Is Not Necessary for Perfect Discrimination between Unitary Operations}
\author{Runyao Duan}
\email{dry@tsinghua.edu.cn}
\author{Yuan Feng}
\email{feng-y@tsinghua.edu.cn}
\author{Mingsheng Ying}
\email{yingmsh@tsinghua.edu.cn}

\affiliation{State Key Laboratory of Intelligent Technology and
Systems, Department of Computer Science and Technology, Tsinghua
University, Beijing, China, 100084}

\date{\today}

\begin{abstract}
We show that a unitary operation (quantum circuit) secretely chosen
from a finite set of unitary operations can be determined with
certainty by sequentially applying only a finite amount of runs of
the unknown circuit. No entanglement or joint quantum operations is
required in our scheme. We further show that our scheme is optimal
in the sense that the number of the runs is minimal when
discriminating only two unitary operations.
\end{abstract}

\pacs{ 03.65.Ta, 03.65.Ud, 03.67.-a}

\maketitle

Entanglement is a valuable physical resource for accomplishing many
useful quantum computing and quantum information processing tasks
\cite{M00}.  For certain tasks such as superdense coding \cite{BS92}
and quantum teleportation \cite{BBC+93}, it has been demonstrated
that entanglement is an indispensable ingredient. For many other
tasks entanglement is also used to enhance the efficiency
\cite{CP00, AC01, DPP01, JFDY06}. One important instance among these
tasks is the discrimination of unitary operations. Although two
nonorthogonal quantum \textit{states} cannot be discriminated with
certainty whenever only finitely many number of copies are available
\cite{CH01}, a perfect discrimination between two different
\textit{unitary} can always be achieved by taking a suitable
entangled state as input and then applying only a finite number of
runs of the unknown unitary operation \cite{AC01,DPP01}. It is
widely believed that this remarkable effect is essentially due to
the use of quantum entanglement. As entanglement is a kind of
nonlocal correlation existing between different quantum systems,
creation of entanglement needs to perform joint quantum operations
on two or more systems. These joint operations are generally
difficult and expensive. Consequently, it is of great importance to
consume as small amount of entanglement as possible in accomplishing
a given task. This motivates us to ask: ``What kind of tasks can be
achieved without entanglement?"

Some pioneering works have been devoted to a good understanding of
the exact role of quantum entanglement in the context of quantum
computing. It has been shown that for certain problems, including
Deutsch-Joza's problem \cite{DJ92}, Simon's problem \cite{SIM97},
and quantum search problem \cite{GRO97}, quantum computing devices
may still have advantages over than any known classical computing
devices even without the presence of entanglement \cite{LL99, ME00,
BBKM04, KMR05}. It was also argued that it may be the interference
and the orthogonality but not the entanglement which are responsible
for the power of quantum computing \cite{ME00}.

In this letter we contribute a new instance of this kind of problems
in the context of quantum information by reporting a somewhat
counterintuitive result: Entanglement is not necessary for perfect
discrimination between unitary operations. We achieve this goal by
explicitly constructing a simple scheme where no entanglement is
needed to discriminate any two given unitary operations with
certainty.

\begin{figure}[ht]
  \centering
  \includegraphics[scale=0.6]{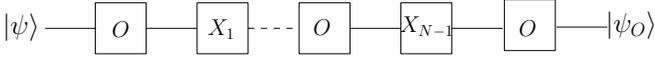}
  \caption{A sequential scheme for discriminating unitary operations $U$ and $V$ without entanglement
   or joint quantum operations. Here $O$ represents the unknown circuit, $N$ is the number of the runs of applying $O$,
   $X_1,\cdots, X_{N-1}$ are the auxiliary unitary
   operations, and $\ket{\psi}$ is the input state. The output states
  $\ket{\psi_U}$ and $\ket{\psi_V}$ are orthogonal.}
  \label{1}
\end{figure}

The basic idea behind our scheme can be best understood in the
following scenario. Suppose we are given an unknown quantum circuit
which is secretely chosen from two alternatives: $U$ or $V$. Here
both $U$ and $V$ are unitary operations acting on a $d$-dimensional
Hilbert space (qudit). To determine which case it really is, we
first apply this circuit to a qudit initially prepared in some state
$\ket{\psi}$. This action will transform the state of the system
into $U\ket{\psi}$ or $V\ket{\psi}$, depending on the unknown
circuit is $U$ or $V$. If there exists a suitable $\ket{\psi}$ such
that the above resulting states are orthogonal, then a perfect
discrimination is achieved. If such a state does not exist, we apply
a suitable unitary operation, say $X_1$, on the above qudit and
apply the unknown circuit once more. After these two runs the state
of the qudit becomes $UX_1U\ket{\psi}$ or $VX_1V\ket{\psi}$.
Similarly, if there exists a suitable input state $\ket{\psi}$ and
unitary operation $X_1$ such that the resulting states are
orthogonal, then a perfect discrimination is achieved again.
Otherwise repeat this procedure. After $N$ runs, the final state is
$\ket{\psi_U}=UX_{N-1}U\cdots X_1U\ket{\psi}$ or
$\ket{\psi_V}=VX_{N-1}V\cdots X_1V\ket{\psi}$. Interestingly, there
always exist a finite $N$, a sequence of unitary operations $X_1,
\cdots, X_{N-1}$, and a suitable input state $\ket{\psi}$ such that
the final output states $\ket{\psi_U}$ and $\ket{\psi_V}$ are
orthogonal. See FIG. \ref{1} for a more intuitive demonstration of
this procedure.

A delicate analysis shows that the number of the runs needed in the
above protocol is equal to that in the original protocol
\cite{AC01,DPP01} and is optimal in any scheme that can perfectly
discriminate $U$ and $V$. It is clear in the above scheme
entanglement is not used. All we need is the ability to perform
unitary operations and projective measurements on a single qudit,
which can be implemented efficiently and economically in experiment.

Let us begin with some preliminaries that are useful in presenting
our main results. We will denote the $d$-dimensional Hilbert space
by $\mathcal{H}_d$.  The notion $\mathcal{U}(d)$ represents the set
of unitary operations acting on $\mathcal{H}_d$. When the dimension
is clear from the context, we omit $d$ for simplicity. For a unitary
operation $U$, we denote by $\Theta(U)$ the length of the smallest
arc containing all the eigenvalues of $U$ on the unit circle. It is
obvious that $\Theta(U)=\Theta(U^\dagger)$ and
$\Theta(U)=\Theta(XUX^\dagger)$ for any $X\in \mathcal{U}$. We say
unitary operations $U$ and $V$ are different if $U$ is not of the
form $e^{i\theta}V$ for any real $\theta$.

Applying the notations introduced above, we can restate the main
ideas in Refs. \cite{AC01,DPP01} as follows. Two unitary operations
$U$ and $V$ are perfectly distinguishable if and only if
$\Theta(U^\dagger V)\geq \pi$. The perfect distinguishablity between
$U$ and $V$ in the multiple-run scenario means there always exists a
finite $N$ such that $\Theta((U^\dagger V)^{\otimes N})\geq \pi$,
which is essentially due to the inequality $\Theta(W^{\otimes
k})\geq\min\{k\Theta(W), \pi\}$ for any unitary $W$ and $k\geq 1$.
The minimal $N$ such that $\Theta((U^\dagger V)^{\otimes N})\geq
\pi$ is given by $\lceil \frac{\pi}{\Theta(U^\dagger V) }\rceil$.
Here $\lceil x \rceil$ denotes the smallest integer that is not less
than $x$. The protocol that discriminates $U$ and $V$ with certainty
consists of three steps: (1) Prepare an $N$-qudit input state
$\ket{\psi}$; (2) Apply the unknown circuit $N$ times on
$\ket{\psi}$ (each qudit one time); (3) Perform a projective
measurement on the output states. Intuitively, this kind of protocol
is called parallel scheme. We should point out that the input state
$\ket{\psi}$ such that $U^{\otimes N}\ket{\psi}$ and $V^{\otimes
N}\ket{\psi}$ are orthogonal should be an $N$-qudit entangled state.
How to generate such an entangled state is a formidable task up to
now even for moderately large $N$. Consequently, this kind of scheme
can be implemented neither efficiently nor economically in practice.

Let us consider a different scheme. We perform the unknown circuit
on the input state step by step. In contrast to the parallel scheme,
this scheme is intuitively named sequential scheme. To enable the
sequential scheme as powerful as possible, we insert a suitable
unitary operation between each two runs of the unknown circuit. This
action can adapt the output state of the previous run to be the best
input state for the next run. Surprisingly, sequential scheme always
leads to a perfect discrimination between any two unitary
operations.

\begin{theorem}\label{noentanglement}\upshape
Let $U$ and $V$ be two different unitary operations, and let
$N=\lceil \frac{\pi}{\Theta(U^\dagger V) }\rceil$. Then there exist
$X_1,\cdots, X_{N-1}\in \mathcal{U}$ and $\ket{\psi}\in \mathcal{H}$
such that
$$UX_{N-1}\cdots X_1U\ket{\psi}\perp VX_{N-1}\cdots X_1V\ket{\psi}.$$
\end{theorem}
{\bf Proof.} For simplicity, we consider first the case where $V$ is
the identity, and then reduce the general case to this special one.
We shall show the following claim: For any nontrivial
$U\in\mathcal{U}(d)$ and $N=\lceil \frac{\pi}{\Theta(U)}\rceil$,
there exists $X\in \mathcal{U}(d)$ such that $\Theta(X^\dagger
UXU^{N-1})\geq \pi$. In other words, there exists a state
$\ket{\psi}\in \mathcal{H}_d$ such that $X\ket{\psi}$ and
$UXU^{N-1}\ket{\psi}$ are orthogonal, and thus $U$ and $I$ are
perfectly distinguishable by $N$ uses.

Let us consider first the case when $d=2$. By the spectral
decomposition theorem, we may assume that $U$ is of the form
$\diag(e^{i\theta}, 1)$, where $\theta=\Theta(U)\in (0,\pi]$. If
$\theta=\pi$, then letting $N=1$ and $X=I_2$, we can directly
verify the validity of the result. Otherwise, let
$$X=\left(%
\begin{array}{cc}
  \cos\alpha & -\sin\alpha\\
  \sin\alpha& \cos\alpha \\
\end{array}%
\right)$$ be a real rotation, where  $0\leq \alpha\leq
\frac{\pi}{2}$. First we seek $\alpha$ such that $\tr(X^\dagger
UXU^{N-1})=0$, which is equivalent to
$$\cos^2\alpha e^{iN\theta}+\sin^2\alpha e^{i(N-1)\theta}+\sin^2\alpha e^{i\theta}+\cos^2\alpha=0.$$
Noticing $(N-1)\theta <\pi\leq N\theta$, we can fulfil the above
equation by taking
$$\alpha=\tan^{-1}\sqrt{-\frac{{\cos (N\theta}/2)}{\cos ({(N-2)\theta}/{2})}}.$$

Second, for the above $\alpha$, let $X^\dagger
UXU^{N-1}=e^{i\beta}\op{\psi_1}{\psi_1}-e^{i\beta}\op{\psi_2}{\psi_2}$
be the spectral decomposition. Choose
$\ket{\psi}=(\ket{\psi_1}+\ket{\psi_2})/{\sqrt{2}}.$ It is easy to
verify that $\braket{\psi}{X^\dagger UX U^{N-1}}{\psi}=0$.

Now for the general case $d>2$.  We can assume without loss of
generality that $U$ is of the form
$\diag(e^{i\theta_1},e^{i\theta_2},\cdots, e^{i\theta_d})$, where
$0\leq \theta_k\leq \Theta(U)<\pi$. In addition, we assume
$\theta_1=\Theta(U)$ and $\theta_2=0$. Then applying the result in
the case of $d=2$, we confirm the existence of $X_{11}\in
\mathcal{U}(2)$ and $\ket{\psi'}\in \mathcal{H}_2$ such that
$\braket{\psi'}{X_{11}^\dagger U_{11}X_{11}
U^{N-1}_{11}}{\psi'}=0$, where $U_{11}=\diag(e^{i\theta_1}, 1)$.
The proof of the claim is completed by setting $X=X_{11}\oplus
I_{d-2}$ and $\ket{\psi}=\ket{\psi'}\oplus 0_{d-2}$.

Let us continue the proof for the general $V$. Setting $U$ and $N$
as $U^\dagger V$ and $\lceil \frac{\pi}{\Theta(U^\dagger
V)}\rceil$, respectively and applying the above claim, we have the
existence of $X\in \mathcal{U}$ and $\ket{\psi'}\in \mathcal{H}$
such that $X\ket{\psi'}$ and $U^\dagger VX(U^\dagger
V)^{N-1}\ket{\psi'}$ are orthogonal. The proof of the theorem is
completed by letting $X_1=X_2=\cdots=X_{N-2}=U^\dagger$,
$X_{N-1}=XU^\dagger$, and $\ket{\psi}=\ket{\psi'}$. \hfill $\square$\\

The above proof also presents an explicit protocol for
discriminating any two unitary operations without entanglement or
joint operations. It is clear that only two different auxiliary
unitary operations, say, $U^\dagger$ and $X$, are required. This
makes the above scheme actually feasible in experiment. It is also
worth noting that the input states leading to perfect discrimination
for different unitary operations are in general not the same.
Interestingly, when only $2\times 2$ unitary operations are under
consideration, any maximally entangled state of the form
$\ket{\Phi}=(\ket{00}+\ket{11})/\sqrt{2}$ is a universal input. This
is mainly due to the simple fact that two $1$-qubit unitary
operations $U$ and $V$ are perfectly distinguishable if and only if
$\tr(U^\dagger V)=0$, which is also equivalent to $(I\otimes
U)\ket{\Phi}\perp (I\otimes V)\ket{\Phi}$ \cite{AC01}. Of course,
any such input state independent scheme needs to consume a maximally
entangled state.

\begin{figure}[ht]
  \centering
  \includegraphics[scale=0.7]{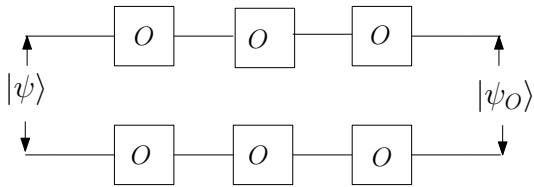}
  \caption{A mixed scheme for discriminating $U$ and $I$. Here $O$
  represents the unknown circuit, $\ket{\psi}$ is a two-qudit input state, and the output state
  $\ket{\psi_O}=(O^{3}\otimes O^{3})\ket{\psi}$.}
  \label{2}
\end{figure}

Combining the parallel scheme with the sequential scheme, we can
design many different mixed schemes for discriminating unitary
operations $U$ and $V$. For simplicity, let us assume $V=I_d$. Let
$1<m<N$, and let $k_1,\cdots, k_m$ be an $m$-partition of $N$, i.e.,
$\sum_{i=1}^m k_i=N$, $k_i\geq 1$. It is clear that discriminating
$U^{k_1}\otimes\cdots\otimes U^{k_m}$ and $I_{d^m}$ with certainty
is sufficient for discriminating $U$ and $I_d$. A simple mixed
scheme is to prepare an $m$-qudit system and then for each $1\leq
i\leq m$ apply $k_i$ times of the unknown circuit to the $i^{th}$
qudit sequentially. FIG. \ref{2} is a mixed scheme with $N=6$,
$m=2$, and $k_1=k_2=3$.

The validity of the scheme is essentially due to the following
inequality
$$\Theta(U^{k_1}\otimes\cdots\otimes
U^{k_m})\geq\min \{(\sum_{i=1}^m k_i)\Theta(U), \pi\}\geq \pi,$$
which can be directly verified by the definition of function
$\Theta$. Any different $m$-partition of $N$ will yield different
mixed scheme. We define the length of the mixed scheme related to
the partition $\{k_i\}$ as $\max_{1\leq i\leq m} k_i$. In practice
we hope the length of the scheme is as small as possible. It is not
difficult to see that the minimal length can be achieved when the
$m$-partition of $N$ is as uniform as possible. The minimal length
is given by $n_{min}= \left\lceil\frac{N}{m}\right\rceil$. Let
$N=(n_{min}-1)m+r$ for some $1\leq r\leq m$. Then a corresponding
partition is $k_1=\cdots= k_r=n_{min}$,
$k_{r+1}=\cdots=k_{m}=n_{min}-1$.

Let us give some  remarks about different schemes for
discrimination. The most advantage of the sequential scheme is that
no entanglement or joint quantum operations is needed. However, any
such kind of scheme needs to perform sequentially at least $N$ times
of the unknown circuit. Instead, in the parallel scheme one needs to
prepare an $N$-partite entangled state as probe state (Here we
notice that it is possible to discriminate two $N$-partite
orthogonal states $U^{\otimes N}\ket{\psi}$ and $V^{\otimes
N}\ket{\psi}$ by using local operations on each single qudit and
classical communications between different qudits only
\cite{WSHV00}, so the measurement device does not require joint
quantum operations). When there are at least $N$ copies of the
unknown circuit and suitable entanglement, we can complete the
discrimination within a single step by applying $N$ copies of the
unknown circuit to the input state simultaneously. For the case when
only $1<m<N$ copies of the unknown circuit are available, the
discrimination task can be finished in $\lceil\frac{N}{m}\rceil$
steps. This reveals an interesting tradeoff between the spatial
resources (entanglement or circuits) and the temporal resources
(running steps or discriminating time). One should choose the most
economic scheme in order to save the resources which are crucial in
practice.

We notice that in the above schemes both the input state
$\ket{\psi}$ and the measurement device for discriminating the final
output states $\ket{\psi_U}$ and $\ket{\psi_V}$ are determined by
$U$ and $V$. When no \textit{a priori} classical information about
the unknown circuit is available, the task is reduced to quantum
operation estimation and it is never possible to achieve a perfect
identification when only finitely many runs (copies) of the unknown
circuit are allowed(available) \cite{AC01}. We would also like to
point out that all the above schemes require the ability of
performing local operations (unitary operations or projective
measurements) on a single qudit in order to perfectly discriminate
the output states. This fact is a little surprising as it seems that
the parallel scheme does not need any auxiliary unitary operations.

For parallel scheme it has been shown that $N=\lceil
\frac{\pi}{\Theta(U^\dagger V) }\rceil$ is the optimal number of the
runs to achieve a perfect discrimination between $U$ and $V$
\cite{AC01}. In what follows we shall prove that this number is also
optimal for perfect discrimination between $U$ and $V$ by using any
sequential scheme. To present this result, we first introduce a key
lemma.
\begin{lemma}\label{chain}\upshape
Let $U$ and $V$ be two unitary operations such that
$\Theta(U)+\Theta(V)<\pi$. Then $\Theta(UV)\leq
\Theta(U)+\Theta(V)$.
\end{lemma}
It is interesting that Lemma \ref{chain} can be directly derived
from Lemma $3$ in Ref. \cite{CP00}. So we omit the proof here.

\begin{theorem}\label{optimal}\upshape
Let $U$ and $V$ be two different unitary operations, and $k<\lceil
\frac{\pi}{\Theta(U^\dagger V) }\rceil$. Then for any unitary
operations $X_1,\cdots, X_{k-1}\in \mathcal{U}$ and $\ket{\psi}\in
\mathcal{H}$,
$$UX_{k-1}\cdots X_1U\ket{\psi}\not\perp VX_{k-1}\cdots X_1V\ket{\psi}.$$
\end{theorem}

{\bf Proof.} Without any loss of generality we may assume that $V=I$
as it is clear that discriminating $U$ and $V$ is equivalent to
discriminating $U^\dagger V$ and the identity $I$.

To prove Theorem \ref{optimal}, it is sufficient to show that if
$k<N=\lceil \frac{\pi}{\Theta(U)}\rceil$ then for any $X_1,\cdots,
X_{k-1}\in \mathcal{U}$, we have
\begin{equation}\label{lesspi}
\Theta((X_{k-1}\cdots X_1)^\dagger (UX_{k-1}\cdots X_1U))<\pi.
\end{equation}
Applying Lemma \ref{chain} $(k-1)$ times we have the following
$$\Theta((X_{k-1}\cdots X_1)^\dagger (UX_{k-1}\cdots X_1U))\leq k\Theta(U),$$
where we have used the fact that $\Theta(X^\dagger UX)=\Theta(U)$
for any unitary $X$. Noticing that $k\leq N-1$ and
$(N-1)\Theta(U)<\pi$,  we have the validity of Eq. (\ref{lesspi}).  \hfill $\square$\\

Employing the similar techniques, we can easily show that $N$ is
also the minimal number of the runs of unknown circuit in any scheme
(sequential scheme, parallel scheme, or any mixed scheme) that
perfectly discriminates $U$ and $V$.

It is straitforward to show that any $n>2$ different unitary
operations can be perfectly distinguishable without entanglement or
joint quantum operations. Let $U_1,\cdots, U_n$ be $n$ possible
candidates, and let $N_{ij}=\lceil\frac{\pi}{\Theta(U_i^\dagger
U_j)}\rceil$, where $1\leq i<j\leq n$. By assuming the circuit is in
$\{U_1,U_2\}$ and then applying the sequential scheme, we can reduce
at least one candidate and only need to consider the left $n-1$
ones.  Repeating this process at most $n-1$ times, we  complete the
discrimination without entanglement or joint operations. The total
number of the runs $N$ satisfies $N_{max}\leq N\leq (n-1)N_{max}$,
where $N_{max}=\max\{N_{i,j}\}$. However, in some special cases the
procedure described above is far from the optimal one. An
interesting example is as follows.

Let $\{\ket{0},\cdots,\ket{d-1}\}$ be an orthonormal basis for
$\mathcal{H}_d$. Consider the set of $d^2$ generalized Pauli
matrices $\{\sigma_{mn}:0\leq m,n\leq d-1\}$, where
$\sigma_{mn}=\sum_{k=0}^{d-1} \omega^{nk}\op{k+m}{k}$ and
$\omega=e^{\frac{2\pi i}{d}}$.  One can readily verify that any
$d\times d$ maximally entangled state can be used to perfectly
discriminate this set of unitary operations. How many runs are
needed if the use of entanglement or joint operations is forbidden?
It is obvious that a single run is not sufficient as there cannot be
$d^2$ orthogonal states in a $d$-dimensional state space. Applying
the procedure described above a $d^2-1$ upper bound can be obtained.
We can do much better by employing a more efficient protocol. Let
$\{\ket{\overline{l}}:0\leq l\leq d-1\}$ be another orthonormal
basis such that
$\ket{\overline{l}}=\frac{1}{\sqrt{d}}\sum_{k=0}^{d-1}\omega^{kl}\ket{k}$.
Then it is easy to see that $(\sigma_{mn}\otimes
\sigma_{mn})\ket{0\overline{0}}=\omega^{-mn}\ket{m\overline{n}}$.
Intuitively, by measuring the first qudit we obtain the index $m$,
while by measuring the second qudit we know the index $n$. Therefore
two runs are necessary and sufficient to discriminate $d^2$ Pauli
matrices without entanglement or joint operations whenever how large
$d$ is. This example also demonstrates that entanglement may reduce
the number of the runs when discriminating $n>2$ unitary operations.

In conclusion, we present a sequential scheme using only unitary
operations and projective measurements to perfectly discriminate
unitary operations. No entanglement or joint quantum operations is
required. This implies that entanglement is not essential in
achieving the perfect discrimination between unitary operations, and
in some sense, confirms the importance of interference and
orthogonality, as suggested in Ref. \cite{ME00}. We also propose
various mixed schemes for discrimination and show the optimality of
these schemes. Notably, there exists an interesting tradeoff between
the spatial resources and the temporal resources. These results
would be helpful when we try to achieve a perfect discrimination
with the lowest cost.

We thank Tal Mor, Zhengfeng Ji, and Guoming Wang for inspiring
suggestions and helpful comments. The acknowledgement is also given
to the other colleagues in the Quantum Computation and Quantum
Information Research Group for enjoyable conversations. This work
was partly supported by the Natural Science Foundation of China
(Grant Nos. 60621062, 60503001, and 60433050), the Tsinghua Basic
Research Foundation (Grant Nos. 052220204 and 052420003), and the
Hi-Tech Research and Development Program of China (Grant No.
2006AA01Z102).

\end{document}